\documentclass[iop]{emulateapj}

\shorttitle{Extended narrow-line emission in HE~2211-3903}
\shortauthors{Julia Scharw\"achter, Michael A. Dopita, Jens Zuther et al.}

\begin{document}


\title{Extended narrow-line emission in the bright Seyfert 1.5 galaxy HE~2211-3903}


\author{J. Scharw\"achter\altaffilmark{1}, M. A. Dopita\altaffilmark{1}, J. Zuther\altaffilmark{2}, S. Fischer\altaffilmark{2}, S. Komossa\altaffilmark{3,4,5,6}, and A. Eckart\altaffilmark{2}}


\altaffiltext{1}{Research School of Astronomy and Astrophysics, The Australian National University, Mount Stromlo Observatory, Cotter Road, Weston Creek 2611, Australia; julia@mso.anu.edu}
\altaffiltext{2}{I. Physikalisches Institut, Universit\"at zu K\"oln, Z\"ulpicher Stra\ss e 77, 50937 K\"oln, Germany}
\altaffiltext{3}{Max-Planck-Institut f\"ur extraterrestrische Physik, Postfach 1312, 85741 Garching, Germany}
\altaffiltext{4}{Technische Universit\"at M\"unchen, Fakult\"at f\"ur Physik, Lehrstuhl f\"ur Physik I,
James Franck Stra\ss e 1/I, 85748 Garching, Germany}
\altaffiltext{5}{Excellence Cluster Universe, TUM, Boltzmannstra\ss e 2, 85748 Garching, Germany}
\altaffiltext{6}{Max Planck Institut f\"ur Plasmaphysik, Boltzmannstra\ss e 2, 85748 Garching, Germany}


\begin{abstract}
Extended narrow-line regions (ENLRs) and extended emission-line regions (EELRs) have been the focus of integral field spectroscopy aiming at the inner kiloparsecs of nearby Seyfert galaxies as well as the larger environment of high redshift QSOs. Based on observations with the Wide Field Spectrograph WiFeS at the 2.3~m telescope of the Australian National University, we present spatially resolved emission-line diagnostics of the bright Seyfert 1.5 galaxy \object{HE~2211-3903} which is drawn from a sample of the brightest Seyfert galaxies at $z<0.06$ with luminosities around the classical Seyfert/QSO demarcation. In addition to the previously known spiral arms of \object{HE~2211-3903}, the emission-line maps reveal a large scale ring with a radius of about 6~kpc which is connected to the active galactic nucleus (AGN) through a bar-like structure. The overall gas kinematics indicates a disk rotation pattern. The emission-line ratios show Seyfert-type, \ion{H}{2} region-type, and composite classifications, while there is no strong evidence of LINER-type ratios. Shock ionization is likely to be negligible throughout the galaxy. The composite line ratios are explained via a mixing line between AGN and \ion{H}{2} region photoionization. Composite line ratios are predominantly found in between the \ion{H}{2} regions in the circum-nuclear region, the bar-like structure to the east of the nucleus, and the eastern half of the ring, suggesting AGN photoionization of the low-density interstellar medium in an ENLR on galaxy scales. The line ratios in the nucleus indicate N-enrichment, which is discussed in terms of chemical enrichment by Wolf-Rayet and Asymptotic Giant Branch stars during past and ongoing nuclear starburst activity.

\end{abstract}





\section{Introduction}

Emission-line diagnostics has been widely applied to emission-line galaxies as a tool for classifying AGN and for separating the effects of the AGN on the interstellar medium in the galaxy from photoionization by \ion{H}{2} regions \citep[e.g.][]{1981PASP...93....5B, 1987ApJS...63..295V, 2006MNRAS.372..961K}. While many of these studies are based on single data points per galaxy, spatially resolved emission-line diagnostics of individual active galaxies has become feasible with modern observational techniques. \citet{2006A&A...446..919B, 2006A&A...456..953B, 2006A&A...459...55B} use long-slit spectra to obtain radial profiles of spatially resolved emission-line diagnostics in Seyfert~2 and Seyfert~1 galaxies. Among other physical parameters, the authors derive a size of the narrow-line region (NLR) based on the change from AGN-type to \ion{H}{2}-type line ratios along the slit. Such a radial profile analysis can be extended to a fully two-dimensional assessment of emission-line diagnostics by means of integral field spectroscopy. 

Several nearby Seyfert galaxies, including the famous examples of Circinus and NGC~1068, have been analyzed using integral field spectrographs with a fine spatial sampling and a typically small field of view. These studies provide insights into nuclear inflows and outflows, excitation, and the connection between AGN activity and nuclear star formation at high physical spatial resolutions, by using adaptive-optics-assisted near-infrared instruments such as SINFONI at the VLT (ESO) \citep[e.g.][]{2006ApJ...646..754D, 2006A&A...454..481S, 2007A&A...466..451Z, 2007ApJ...671.1388D, 2009ApJ...702..114D, 2009ApJ...691..749S, 2009ApJ...698.1852B, 2010A&A...519A..79F} or NIFS at Gemini North \citep[e.g.][]{2006MNRAS.373....2R, 2008MNRAS.385.1129R, 2009MNRAS.393..783R, 2009MNRAS.394.1148S,2010MNRAS.404..166R, 2010MNRAS.402..819S}. Seeing-limited optical integral field spectroscopy of the circum-nuclear region in nearby Seyfert galaxies include observations with GMOS at Gemini \citep[e.g.][]{2006MNRAS.371..170B, 2009MNRAS.396....2B}, SAURON at the 4.2-m William Herschel Telescope \citep[e.g.][]{2005MNRAS.364..773F, 2006MNRAS.365..367E, 2007MNRAS.379.1249D}, or OASIS at the CFHT \citep{2009A&A...500.1287S}. This type of observations probes the NLR gas in the inner $\lesssim 1$~kpc of the host galaxy around the moderate-luminosity AGN in nearby Seyfert galaxies.

The AGN, however, can influence the surrounding interstellar medium on much larger scales. Evidence for large-scale AGN ionization is found in the form of extended narrow-line regions (ENLRs) on scales of up to $\sim 20$~kpc in Seyfert galaxies and extended emission-line regions (EELRs) of tens of kpc around higher redshift QSOs \citep[e.g.][]{1987MNRAS.228..671U, 1987ApJ...316..584S}.
EELRs and AGN feedback in QSOs and radio galaxies at intermediate to high redshift has become another focus of multi-slit and, ultimately, integral field spectroscopy \citep[e.g.][]{2010IAUS..267..429M}. \citet{2008A&A...491..407N} find EELRs on scales of up to 30~kpc around three powerful radio galaxies at a redshift of 2 to 3, indicative of radio-jet-driven outflows. \citet{2009ApJ...690..953F} analyze the emission-line diagnostics of EELRs around mostly radio-loud $z\sim 0.1-0.4$ quasars and discuss these in a scenario of quasar superwinds. The EELRS as well as the broad-line regions of their central quasars show low metallicities. \citet{2008A&A...488..145H,2010A&A...519A.115H} find EELRs around a mostly radio-quiet sample of $z< 0.3$ QSOs from the Palomar-Green and Hamburg/ESO Surveys based on integral field spectroscopy with the Potsdam Multi-Aperture Spectrophotometer PMAS at the 3.5~m telescope at Calar Alto. 

In order to fill the transition region between the detailed nuclear studies of moderate-luminosity nearby Seyfert galaxies, on the one hand, and the global studies of ionized gas around bright QSOs and radio galaxies at intermediate to high redshift, on the other hand, we focus on two-dimensional emission-line diagnostics of some of the brightest AGN at $z<0.06$. The project is based on a sample of 99 type-1 AGN selected from the Hamburg/ESO bright QSO sample \citep{2000A&A...358...77W} according to a redshift criterion $z<0.06$, as described in \citep{2007A&A...470..571B}. This sample of borderline type-1 QSOs comprises bright local AGN, clustering around the classical Seyfert/QSO demarcation magnitude \citep[see][]{1997A&A...325..502K}. Based on millimeter observations, \citet{2007A&A...470..571B} suggest that the sample objects show transition properties between the lower-luminosity Seyfert 1 galaxies and luminous QSOs in terms of the far-infrared to CO luminosity. Parts of the sample have been the focus of \ion{H}{1} observations as well as near-infrared long-slit spectroscopy \citep{2006A&A...452..827F, 2009A&A...507..757K}.
 
In this paper, we present a host-galaxy-wide analysis of ionized gas for the bright Seyfert galaxy \object{HE~2211-3903} (\object{ESO344-G016}) at $z=0.0397$, drawn from the sample of borderline type-1 QSOs. Using integral field spectroscopy, we map emission-line diagnostics throughout the host galaxy of \object{HE~2211-3903} in order to separate ionization by the AGN from ionization by star formation. The properties of the AGN continuum source, gas-phase metallicities, and AGN-photoionized extended line emission are assessed by comparing the emission-line ratios with available photoionization models. 
\object{HE~2211-3903} is a Seyfert 1.5 galaxy \citep{2006A&A...455..773V} hosted in a spiral galaxy which has been cited as Sb galaxy \citep{1982euse.book.....L} and, in the more recent literature, as barred galaxy \citep{2003AJ....126.1750M, 2006A&A...452..827F}. It is associated with an IRAS source and belongs to the class of luminous infrared galaxies. The infrared luminosity is $L_{IR} = 1.4\times 10^{11}\ L_\odot$, using the definition from \citet{1996ARA&A..34..749S}. Our data present the first detailed case study of \object{HE~2211-3903}. $H_0 = 71\ \mathrm{km\ s^{-1}\ Mpc^{-1}}$, $\Omega_M = 0.27$, and $\Omega_{vac} = 0.73$ is assumed throughout the paper, resulting in a physical scale of 0.78~kpc~arcsec$^{-1}$ for HE~2211-3903 \citep{2006PASP..118.1711W}.

\section{Observations and Data Reduction}\label{sec:datared}

\begin{figure}
\epsscale{1.0}
\plotone{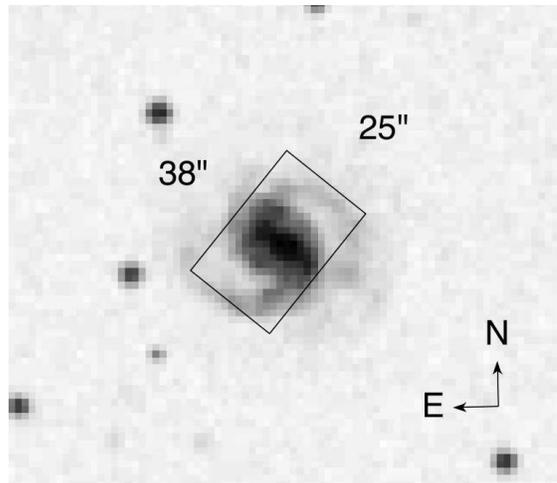}
\caption{SERC-J image of HE~2211-3903 from the Digitized Sky Survey (filter GG395). The overlay shows the $25\arcsec \times 38\arcsec$ field-of-view of WiFeS for the orientation used during the observations.\label{fig:dss}}
\end{figure}

Integral field spectroscopy of \object{HE~2211-3903} was carried out with the 
Wide Field Spectrograph WiFeS at the 2.3~m telescope of the Australian National University
at Siding Spring Observatory in Australia (Fig.~\ref{fig:dss}). 
WiFeS is an image-slicer, providing a field-of-view
of $25\arcsec \times 38\arcsec$ via 25 slitlets of 38\arcsec $\times 1$\arcsec. 
A wide spectral coverage is achieved via simultaneous observations
in a blue and a red arm. The choice of gratings includes spectral resolutions of 3000 and 7000.
A detailed description of the instrument and its characteristics can be found in \citet{2007Ap&SS.310..255D, 2010Ap&SS.327..245D}.

The WiFeS data for \object{HE~2211-3903} were taken in October 2009 using the blue $B3000$ and the
red $R7000$ gratings. The data were obtained in nod-and-shuffle mode using cycles of 100~s integrations on the object and
50~s integrations on the sky. The total observing time on the science object was split into five frames in order to allow for cosmic ray removal via 
median-combination.

The data reduction makes use of the WiFeS IRAF pipeline \citep[see][]{2010Ap&SS.327..245D} with user-specific modifications for bias and sky subtraction.
The bias subtraction is based on a combination of overscan-subtraction and the subtraction of a low-order surface fit to a single bias observed close in time to the object at night.
This method is found to yield optimal results given the curved and time-variable bias levels of the four amplifiers of the WiFeS CCDs during the October 2009 observations. In some cases the bias subtraction is
additionally improved by manual fine-tuning of the surface fit. Residual bias artifacts in the continuum level across the spectrum are treated with a second order correction for the purpose of continuum subtraction
(see Section~\ref{sec:continuum}).
The sky subtraction in nod-and-shuffle mode consists of the subtraction of the sky slices from the corresponding object slices. Because of the unequal integration times on object and sky, the sky slices are scaled by a factor of two.
The associated increase in noise in the sky slices is reduced by a spatial two-pixel boxcar-smoothing.
The red spectra are corrected for telluric absorption using telluric standard star observations of $B1$ to $B4$ type stars. No telluric correction is applied to the blue spectra. The flux calibration is based on an average sensitivity curve derived from two spectrophotometric standard stars. As some of the object frames are affected by passing clouds, the flux calibration is applied to the good frames only while the other frames are scaled accordingly before median-combination.

The WiFeS pipeline provides output data cubes with rectangular $0.5\arcsec \times 1\arcsec$ pixels. Square pixels of $1\arcsec \times 1\arcsec$ at an enhanced S/N are obtained by coadding two pixels of the cube in y-direction.
The spatial resolution of the data is estimated based on the assumption that the broad component of the nuclear H$\alpha$ emission in \object{HE~2211-3903} is intrinsically unresolved. Fitting a circular Gaussian profile to the point-spread function (PSF) of the broad H$\alpha$ emission results in a FWHM of 2\arcsec.

The wings of the AGN PSF as well as scattered light from the AGN are potential contaminations in the off-nuclear emission-line data discussed in this paper. The effects of AGN contamination were assessed by re-analysing the continuum-subtracted data cube (Section~\ref{sec:continuum}) after subtracting a scaled AGN spectrum from all spectral pixels for which a broad Balmer component is detected. The AGN spectrum is assumed to be the spectrum in the brightest spectral pixel of the AGN in the continuum-subtracted cube. The spectrum is spatially scaled according to the spatial flux distribution of the broad Balmer components. Scaling with flux in both, broad H$\alpha$ as well as broad H$\beta$, was tested separately. Besides introducing new uncertainties and additional noise, the AGN subtraction was found to have negligible effects on the results for the off-nuclear emission-line analysis. The data in this paper are presented without AGN subtraction.

\section{Analysis}

\subsection{Continuum subtraction\label{sec:continuum}}

The data cube is continuum-subtracted and cleaned from stellar absorption lines using an adapted version of the Penalized Pixel-Fitting method (pPXF), \citep{2004PASP..116..138C}.
A combination of stellar templates of G and K supergiants from the Indo-U.S. Coud\'e Feed Spectral Library \citep{2004ApJS..152..251V} together with a low-order polynomial are fitted to the combined blue and
red spectra for each spectral pixel individually. Residual discontinuities in the spectra due to bias subtraction artifacts and flux calibration offsets between the blue and red spectra are equalized beforehand. This correction is done by adding/subtracting a constant value determined from the average of the continuum to the left/right of the discontinuity. 
The stellar templates from the Indo-U.S. Coud\'e Feed Spectral Library have a spectral resolution of 1~\AA\ FWHM over the wavelength range from 3460~\AA\  to 9464~\AA. This approximately matches the spectral resolution in the red part of the WiFeS spectrum taken with the $R7000$ grating. The $B3000$ grating used for the blue WiFeS spectrum has a lower spectral resolution. In order to account for this mismatch, the blue part of the stellar template spectra is smoothed to the lower spectral resolution before the fitting process.
As the S/N of the stellar absorption lines strongly varies across the field-of-view, low S/N spectral pixels are binned using a spatial boxcar smoothing. The smoothing box for each spectral pixel is increased iteratively until either a required S/N is reached or the smoothing box exceeds a maximum size of $9\times 9$ spectral pixels. If the required S/N is reached, the stellar templates are fitted to the continuum together with a low-degree polynomial according to the pPXF method. If the smoothing box limit is exceeded without reaching the required S/N, the continuum of the respective spectral pixel is fitted using a higher-degree polynomial only. 
All fits are performed after blanking out emission lines as well as regions around the corrected continuum discontinuities.

\begin{figure*}
\epsscale{1.0}
\plotone{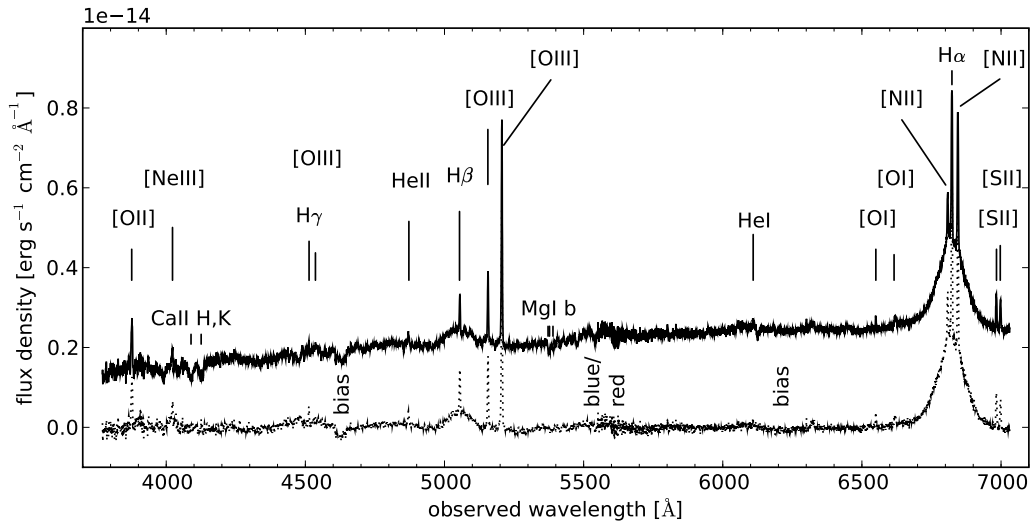}
\caption{Comparison of the nuclear spectrum of HE~2211-3903 before (solid line) and after (dotted line) continuum subtraction. Both spectra show the combined wavelength range of the blue and the red WiFeS spectrum averaged over a 3 pixel (i.e. 3\arcsec) circular aperture. The most dominant emission and absorption lines are marked. ``Bias'' and ``blue/red'' indicate the regions of imperfect bias corrections and the transition from the blue to the red spectrum.\label{fig:stellsub}}
\end{figure*}

The resulting continuum fit is subtracted from the original unbinned data cube in order to retain the maximum spatial resolution for the emission-line studies. A nuclear spectrum before and after subtraction of the continuum and stellar absorption lines is shown in Fig.~\ref{fig:stellsub}.
Those spectral pixels which are dominated by the AGN featureless continuum, can be affected by inaccuracies in the stellar continuum subtraction \citep[see][]{2006A&A...459...55B}. A significant oversubtraction of stellar absorption lines for \object{HE~2211-3903} can be excluded based on the fact that there is virtually no difference in emission-line ratios before and after continuum subtraction.

\subsection{Emission-line fitting\label{sec:linefitting}}

Emission lines are fitted with Gaussian line profiles in an implementation of the IDL routine MPFITPEAK. All lines are fitted with a single Gaussian component plus a zero-order polynomial. Only for the hydrogen lines an additional broad Gaussian component is added in order to account for the AGN broad-line emission in the nuclear PSF. 
The emission lines across the whole spectral range are fitted in two blocks which account for line blending. In the red part of the spectrum [\ion{O}{1}]~$\lambda$6300, [\ion{O}{1}]~$\lambda$6364, [\ion{N}{2}]~$\lambda$6548, H$\alpha$, [\ion{N}{2}]~$\lambda$6583, [\ion{S}{2}]~$\lambda$6717, and [\ion{S}{2}]~$\lambda$6731 are fitted simultaneously. The line kinematics in terms of line centers and widths is tied to the narrow H$\alpha$ component.
In the blue part of the spectrum H$\beta$, [\ion{O}{3}]~$\lambda$4959, and [\ion{O}{3}]~$\lambda$5007 as well as [\ion{O}{2}]~$\lambda$3727 and [\ion{O}{2}]~$\lambda$3729 are fitted simultaneously. Here the line kinematics
 is tied to H$\beta$. [\ion{O}{2}]~$\lambda$3727 and [\ion{O}{2}]~$\lambda$3729 are included in this fit despite their large separation in wavelength, as tying their kinematics to H$\beta$ provides better constraints on their fit in this lower S/N part of the blue spectrum.
The broad hydrogen components of H$\alpha$ and H$\beta$ are removed from the fit if their line widths exceed a specified limit. This ensures that a broad-component fit is only attempted where a broad component with a reasonable line width is present, i.e. ideally in the nuclear region only.

The errors of the fit parameters of the emission-line fitting are used for error propagation and data clipping in the further analysis.  These errors, provided by MPFITPEAK, are computed from the covariance matrix of the least-squares fit and depend on the estimated uncertainties of the input spectrum. The input uncertainties are based on the standard deviation of the input spectrum in a mostly line-free spectral region. In the fit to the red spectrum, the standard deviation is assumed as a constant error. As the fit to the blue spectrum involves a larger wavelength range, two different constant uncertainties are assigned to the short-wavelength and the long-wavelength part. These are based on the standard deviation in a sample region close to [\ion{O}{2}]~$\lambda$3727 and [\ion{O}{2}]~$\lambda$3729 and close to H$\beta$, [\ion{O}{3}]~$\lambda$4959, and [\ion{O}{3}]~$\lambda$5007, respectively. The resulting reduced chi-squares of the fits for all spectral pixels are of the order of 1. The errors of the fit parameters are, therefore, scaled with the square root of the reduced chi-square. In the following all data points with a relative error larger than 30\% are clipped.

\section{Results}

\subsection{Emission-line maps and kinematics}
\begin{figure*}
\epsscale{0.8}
\plotone{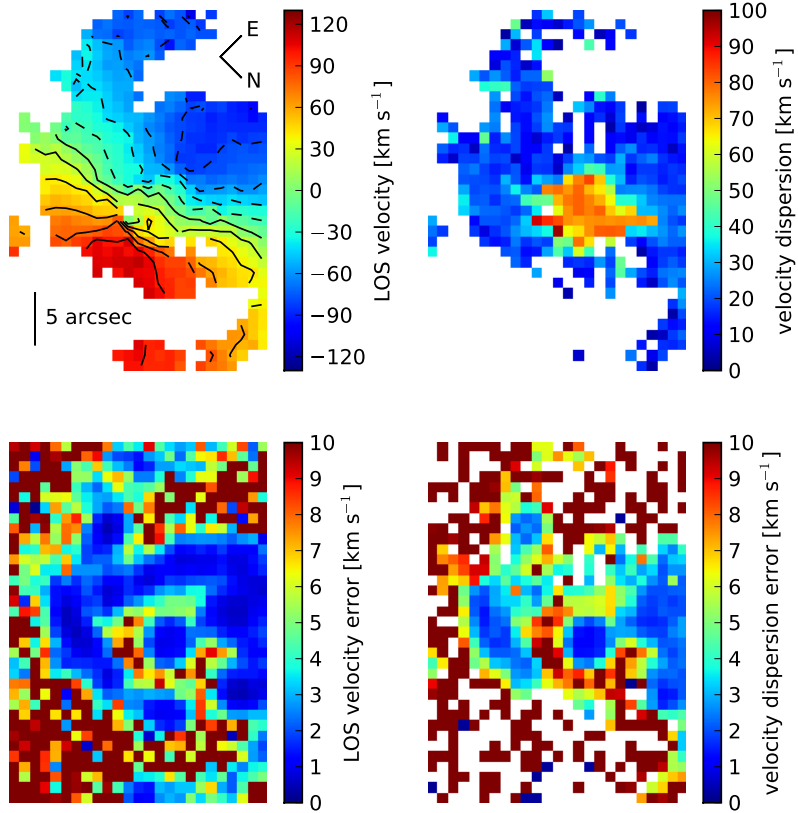}
\caption{Line-of-sight (LOS) velocity and velocity dispersion in HE~2211-3903 (top panels) tied to Gaussian profile fits to the narrow H$\alpha$ component. The LOS velocity is marked with contours from $-100\ \mathrm{km\ s^{-1}}$ to $+100\ \mathrm{km\ s^{-1}}$ in steps of $20\ \mathrm{km\ s^{-1}}$. The bottom panels show the absolute errors for non-zero values of the LOS velocity and velocity dispersion, respectively. Clipped data points are shown in white. The velocity dispersion is corrected for the spatially variable instrumental dispersion of the WiFeS R7000 grating determined from the line width of the 6300~\AA\ sky line.\label{fig:kinematics}}
\end{figure*}

\begin{figure*}
\epsscale{0.8}
\plotone{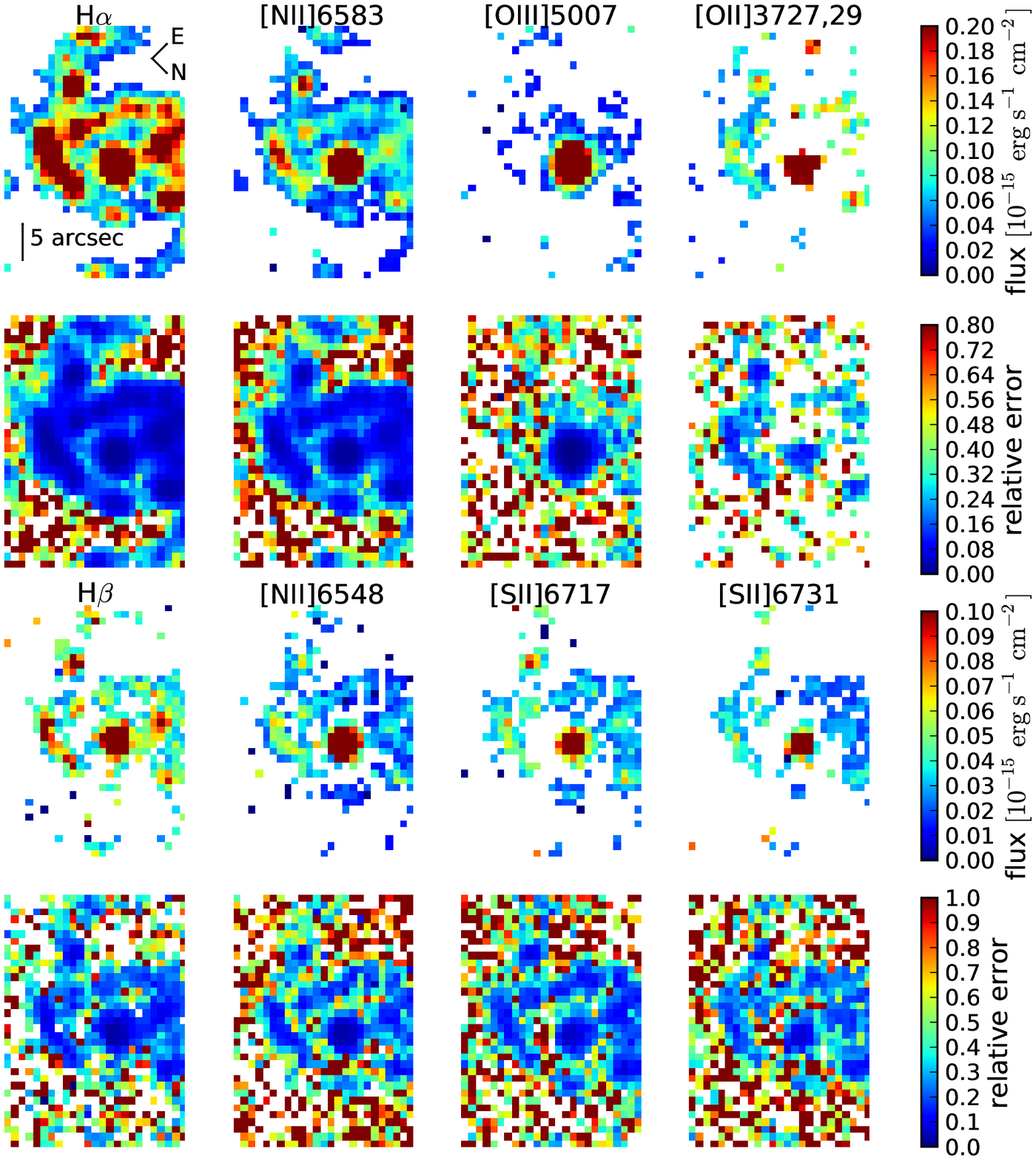}
\caption{Emission-line maps of the main emission lines in HE~2211-3903 (first and third row) and their corresponding relative errors (second and fourth row). For H$\alpha$ and H$\beta$ this map shows the narrow component of the line fit only. The maps and their corresponding errors are derived from the Gaussian line fit as described in Section~\ref{sec:linefitting}. All pixels with a relative error larger than 30\% are clipped. The error maps show relative errors for all pixels with a non-zero line flux. Clipped values are shown in white. See the electronic edition of this Journal for a color version of this figure.\label{fig:linemaps}}
\end{figure*}

The morphology and kinematics of the narrow-component emission-line gas in \object{HE~2211-3903} reveals a rotating disk with a large-scale ring which is connected to the nucleus through a linear structure.
Fig.~\ref{fig:linemaps} shows the emission-line maps for \object{HE~2211-3903} in the narrow components of H$\alpha$ and H$\beta$ as well as in [\ion{N}{2}]~$\lambda$6583, [\ion{O}{3}]~$\lambda$5007, [\ion{O}{2}]~$\lambda$3727,29, [\ion{N}{2}]~$\lambda$6548, [\ion{S}{2}]~$\lambda$6717, and [\ion{S}{2}]~$\lambda$6731. 
H$\alpha$ and other lines tracing star formation show \ion{H}{2} region complexes in the ring in addition to the two spiral arms emanating outwards in the WiFeS field-of-view. The ring has a radius of about 8\arcsec, i.e. 6~kpc. Assuming that the intrinsic shape of the ring is close to circular, the observed ring shape is in agreement with a close to face-on view. An estimate of the inclination is $i \lesssim 35\deg$, which is, however, complicated by ambiguities in the exact location of the \ion{H}{2} regions with respect to the ring and spiral arms.
The major kinematic axis of the global rotation of \object{HE~2211-3903} is close to the east-west direction with a position angle of about $95\deg$ (left panels in Fig.~\ref{fig:kinematics}). The maximum line-of-sight velocities observed are $v_{LOS} \sim \pm 100\ \mathrm{km\ s^{-1}}$. Assuming an inclination of 
$i \lesssim 35\deg$, the maximum circular velocity is $v_{LOS}/\sin(i) \gtrsim  \pm 175\ \mathrm{km\ s^{-1}}$, which is in the typical range of maximum velocities for spiral galaxies \citep[e.g.][]{2005MNRAS.362..127G}.

The ring of ionized gas is connected to the nuclear line emission through a linear structure. The latter is likely to be associated with the stellar bar mentioned in previous publications \citep{2003AJ....126.1750M, 2006A&A...452..827F}. While there is some evidence of diffuse emission from ionized gas along the linear structure, bright \ion{H}{2} regions are only found near the intersection of the linear structure with the ring. Based on this first-order morphological assessment, \object{HE~2211-3903} resembles other examples of ringed galaxies, where the ring is likely to be caused by a bar resonance \citep[e.g.][]{1996ApJS..105..353C}.

The velocity dispersion of the narrow line components in the red spectrum (right panels in Fig.~\ref{fig:kinematics}) shows a clear separation between broadened lines in the nuclear region and narrow lines in the disk. The gas in the ring and spiral arms has velocity dispersions of about 20~$\mathrm{km\ s^{-1}}$, typical of extra-galactic \ion{H}{2} regions \citep[e.g.][and references therein]{1990ARA&A..28..525S}. The broadened lines in the nuclear region trace the AGN NLR. The WiFeS data suggest a velocity dispersion of about 80~$\mathrm{km\ s^{-1}}$ for the kinematics of the gas in the NLR.

A detailed modeling of the kinematics and potential of \object{HE~2211-3903} would exceed the scope of this paper and is left to future studies.

\subsection{Interstellar extinction\label{sec:extinction}}

\begin{figure*}
\epsscale{0.8}
\plotone{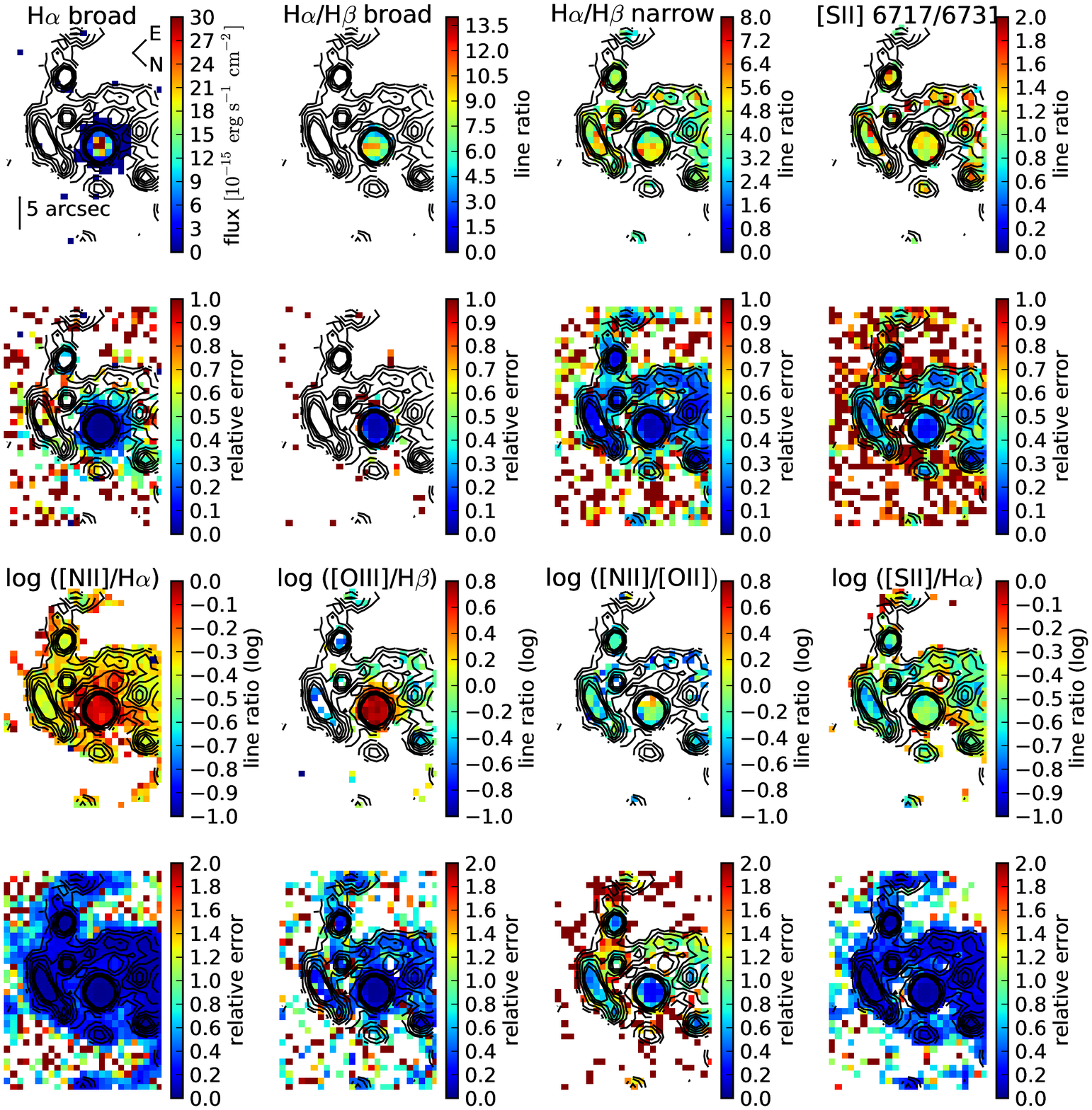}
\caption{Maps of the broad H$\alpha$ component (first row, left panel), the flux ratios of the broad and narrow H$\alpha$ and H$\beta$ components (first row, middle panels), the flux ratio of [\ion{S}{2}]~$\lambda$6717/[\ion{S}{2}]~$\lambda$6731 (first row, right panel), and the logarithm of the flux ratios of [\ion{N}{2}]~$\lambda$6583/H$\alpha$, [\ion{O}{3}]~$\lambda$5007/H$\beta$, [\ion{N}{2}]~$\lambda$6583/[\ion{O}{2}]~$\lambda$3727,29, and  [\ion{S}{2}]~$\lambda$6717,6731/H$\alpha$ (third row). The respective relative errors of the non-zero pixels in the maps are plotted in the second and fourth rows. Clipped data points are displayed in white.\label{fig:allratios}}
\end{figure*}
The Balmer decrement H$\alpha$/H$\beta$ is used to assess and correct for interstellar extinction in the WiFeS spectra of \object{HE~2211-3903}. The third panel in the upper row of Fig.~\ref{fig:allratios} shows the Balmer decrement based on the narrow components of H$\alpha$ and H$\beta$. The values of H$\alpha$/H$\beta$ mostly range between 3 and 5 (i.e. $A_V \approx 0.1$~mag to 1.5~mag). An increase in reddening is observed in the western part of the nucleus and in the large \ion{H}{2} region to the south-west. The flux ratio there reaches values of up to 6 (i.e. $A_V \approx 2$~mag). 
For completeness, the ratio of the broad H$\alpha$ and H$\beta$ components is plotted in the second panel in the upper row of Fig.~\ref{fig:allratios}. In the nuclear PSF, the mean value of the ratio is about 7.
In how far this ratio can be attributed to dust extinction in the BLR is unclear.
The high-density region of the BLR does not necessarily follow pure case 
B recombination \citep{2006agna.book.....O}, since collisional and 
optical depth effects might also play a role. On the other hand, \citet{2008MNRAS.383..581D}
have shown that the Balmer decrements of blue quasars
are indeed in a narrow range close to the recombination value of 3. When using case B recombination as a reference for the broad-line region,
the corresponding visual extinction is $\sim 2.5$~mag to $\sim 3.5$~mag. 

The visual extinction $A_V$ mentioned above is calculated from  
\begin{equation}
A_V = -2.5 \log \left( \frac{F_{H\alpha}/F_{H\beta}}{2.85}\right) \left(\frac{\kappa_{5500\AA}}{\kappa_{H\alpha}-\kappa_{H\beta}}\right),
\end{equation}
where $F$ is the line flux, 2.85 is the theoretical ratio of $F_{H\alpha_0}/F_{H\beta_0}$ for case B recombination \citep{2006agna.book.....O}, and $\kappa$ is the attenuation curve including contributions from absorption and scattering \citep{2003ApJ...599L..21F, 2005ApJ...619..340F}. \citet{1997MNRAS.292..273W} report a nuclear reddening of $E(B-V)=0.63$ for \object{HE~2211-3903}, which corresponds to $A_V = R_V \times E(B-V) = 2.0$ for $R_V = 3.1$.
In the following analysis, extinction corrections are only applied to line fluxes if they are used in line ratios between lines with a large separation in wavelength space. The extinction correction of a line at central wavelength $\lambda$ is based on the relation
\begin{equation}
F_{\lambda_\mathrm{corr}} = F_\lambda \left( \frac{F_{H\alpha}/F_{H\beta}}{2.85}\right) ^{-\kappa_\lambda/(\kappa_{H\alpha}-\kappa_{H\beta})}.
\end{equation}
For simplicity, a constant intrinsic H$\alpha$/H$\beta$ ratio of 2.85, typical of \ion{H}{2} regions, is used throughout the host galaxy. AGN-dominated regions might have a larger intrinsic H$\alpha$/H$\beta$ ratio of 3.1 \citep{2006MNRAS.372..961K}. The difference between the two values implies an uncertainty of about 30\% with some variation depending on the wavelength at which the extinction is assessed. This is a similar order of magnitude as that of other uncertainties in the data.

\subsection{Intrinsic AGN power\label{sec:AGNpower}}

\object{HE~2211-3903} belongs to the bright end of the local Seyfert galaxy luminosity function. With $L_{\mathrm{OIII}} \approx 4\times 10^{41}\ \mathrm{erg\ s^{-1}}\approx 9\times 10^7\ L_\odot$, the [\ion{O}{3}] luminosity of \object{HE~2211-3903}  is close to $L^\star $ of the [\ion{O}{3}] luminosity function of $0 < z < 0.15$ AGN from the Sloan Digital Sky Survey \citep{2005AJ....129.1795H}. Compared to the Seyfert~1 galaxy sample from the latter study \object{HE~2211-3903} shows an [\ion{O}{3}] luminosity which is between the Schechter function $L^\star $ and the one- or two-power-law function $L^\star $ \citep[Table~1]{2005AJ....129.1795H}.

The estimate for the [\ion{O}{3}]~$\lambda$5007 luminosity of \object{HE~2211-3903} is derived from the total [\ion{O}{3}] flux in a nuclear aperture with a diameter of 6\arcsec\ corresponding to three times the FWHM of the Gaussian fit to the PSF of the nuclear broad H$\alpha$ emission (see Section~\ref{sec:datared}). The [\ion{O}{3}] flux, corrected for extinction based on the mean Balmer decrement in the same aperture, is of the order of $F_{\mathrm{[OIII]}} = 1\times 10^{-13}\ \mathrm{erg\ s^{-1}\ cm^{-2}}$. Using a luminosity distance of $D_L = 173\ \mathrm{Mpc}$, the nuclear flux corresponds to a total [\ion{O}{3}]~$\lambda$5007 luminosity $L_{\mathrm{[OIII]}} = 4\pi D_L^2 F_{\mathrm{[OIII]}}$ of the order of $ 4\times 10^{41}\ \mathrm{erg\ s^{-1}}\approx 9\times 10^7\ L_\odot$. \citet{1992MNRAS.257..677W} report an [\ion{O}{3}] luminosity of $\log\ L_{\mathrm{[OIII]}} = 41.39$ in erg~s$^{-1}$, based on long-slit spectroscopy and assuming a Hubble constant of $H_0 = 50\ \mathrm{km\ s^{-1}\ Mpc^{-1}}$. This is the same order of magnitude as the WiFeS measurement of $\log\ L_{\mathrm{[OIII]}} = 41.6\ \mathrm{[erg\ s^{-1}]}$ or $\log\ L_{\mathrm{[OIII]}} = 41.9\ \mathrm{[erg\ s^{-1}]}$ for $H_0 = 75\ \mathrm{km\ s^{-1}\ Mpc^{-1}}$ or $H_0 = 50\ \mathrm{km\ s^{-1}\ Mpc^{-1}}$, respectively. The intrinsic power of the AGN is estimated from the [\ion{O}{3}] luminosity using the bolometric correction of 3200 \citep{2010arXiv1006.5178S}. This results in a bolometric luminosity of the order of $L_{bol} =1\times 10^{45}\ \mathrm{erg\ s^{-1}} = 3\times 10^{11}\ L_\odot$. 
An alternative proxy for the bolometric luminosity of the AGN can be obtained from the 5100~\AA\ luminosity. In a nuclear aperture with a diameter of 6\arcsec\ the extinction-corrected flux density of \object{HE~2211-3903} at 5100~\AA\ is of the order of $f_{\lambda} = 8\times 10^{-15}\ \mathrm{erg\ s^{-1}\ cm^{-2}\ \AA^{-1}}$, i.e. $F_{5100} = \lambda f_{\lambda} = 4\times 10^{-11}\ \mathrm{erg\ s^{-1}\ cm^{-2}}$.  The corresponding luminosity is of the order of $L_{5100} =1\times 10^{44}\ \mathrm{erg\ s^{-1}}$. Using a bolometric correction of 10.3 \citep{2006ApJS..166..470R}, the derived bolometric luminosity of the order of $L_{bol} =1\times 10^{45}\ \mathrm{erg\ s^{-1}}$ is in agreement with the [\ion{O}{3}]-based value.

\object{HE~2211-3903} has been detected in X-rays during the ROSAT all-sky survey \citep{1999yCat.9010....0V} with a PSPC count rate of $0.44 \pm 0.04\ \mathrm{cts\ s^{-1}}$. Assuming only Galactic absorption towards the nucleus of \object{HE~2211-3903} and an intrinsic power-law spectral index of $\Gamma_{\rm x}=-1.9$ results in a 0.1-2.4 keV X-ray luminosity of $L_{\rm x} = 6 \times10^{43}\ \mathrm{erg\ s^{-1}}$. If the X-ray emission is absorbed by an extra X-ray column density of $3.7\times 10^{21}\ \mathrm{cm^{-2}}$ to $6.7\times 10^{21}\ \mathrm{cm^{-2}}$, corresponding to the range of nuclear visual extinction of A$_{V} \approx 2.0$ to A$_{V} \approx 3.6$ (see Section~\ref{sec:extinction}), then the intrinsic luminosity is $L_{\rm x} \approx 3\times 10^{44}\ \mathrm{erg\ s^{-1}}$  to $L_{\rm x} \approx 4.4\times 10^{44}\ \mathrm{erg\ s^{-1}}$. A bolometric correction factor of roughly 10 again yields a bolometric luminosity of the order of $L_{bol} \approx 10^{45}\ \mathrm{erg\ s^{-1}}$.

\subsection{Electron density\label{sec:electrondensity}}
The ratio of  [\ion{S}{2}]~$\lambda$6717/[\ion{S}{2}]~$\lambda$6731, as an indicator of the average electron density \citep[e.g.][]{2006agna.book.....O}, is plotted in the upper right panel in Fig.~\ref{fig:allratios}. 
The average flux ratio of  [\ion{S}{2}]~$\lambda$6717/[\ion{S}{2}]~$\lambda$6731 throughout \object{HE~2211-3903} is about 1.2 to 1.3, corresponding to an electron density of $n_e \approx 100\ \mathrm{cm^{-3}}$ to $300\ \mathrm{cm^{-3}}$ \citep{1993ApJS...88..329C}. The scatter in the ratio from pixel to pixel suggests densities ranging from about $10\ \mathrm{cm^{-3}}$ to about $10^{3.2}\ \mathrm{cm^{-3}}$. In part, this scatter is larger than the error bars inferred from the spectral fitting procedure. If the error bars are not underestimated, the pixel-to-pixel variations might be an indication of $\sim$kpc-scale density variations. The likely presence of small-scale density inhomogeneities in the NLR has been suggested previously \citep[e.g.][]{1997A&A...323...31K}.
\begin{figure*}
\epsscale{0.6}
\plotone{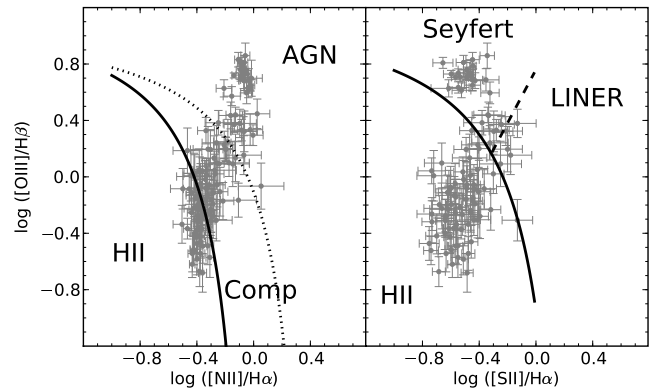}
\caption{Diagnostic diagrams for HE~2211-3903, based on the pixel values and errors from the third and fourth rows in Fig.~\ref{fig:allratios} (gray error bars). The solid, dotted, and dashed lines indicate the borders between the loci of line ratios classified as ``\ion{H}{2}'', ``composite'', ``Seyfert'', ``LINER'', or ``AGN'' (including Seyferts and LINERs), based on \citet{2006MNRAS.372..961K}. An extinction correction is not applied, as the line ratios are computed from lines which are close to each other in wavelength space.\label{fig:bpt}}
\end{figure*}
\begin{figure*}
\epsscale{0.8}
\plotone{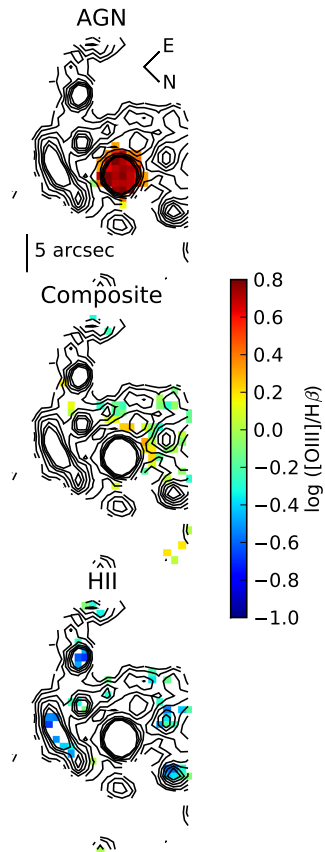}
\caption{Spatial location of pixels classified as AGN, composite, or \ion{H}{2} according to the left panel in Fig.~\ref{fig:bpt} highlighted on the basis of the [\ion{O}{3}]~$\lambda$5007/H$\beta$ map from Fig.~\ref{fig:allratios}. The contours from the flux distribution of the narrow H$\alpha$ line are overplotted. See the electronic edition of the Journal for a color version of this figure including color-coded pixel values.\label{fig:compositeAGNHII}}
\end{figure*}

\subsection{Emission-line diagnostics\label{sec:diagnosticratios}}

\object{HE~2211-3903} shows evidence of Seyfert-type, \ion{H}{2}-region-type, and composite line ratios, while there is no clear indication of LINER-type emission-line gas.
The spatial variation of the common diagnostic ratios throughout \object{HE~2211-3903} together with the relative error for each pixel are displayed in the third row of Fig.~\ref{fig:allratios}. 
The maps of [\ion{N}{2}]/H$\alpha$ and [\ion{O}{3}]/H$\beta$ in Fig.~\ref{fig:allratios} clearly separate gas ionized by the AGN from gas ionized by young stars in \ion{H}{2} regions. The relation between the two line ratios is plotted together with the ``\ion{H}{2} region'', ``composite'', and ``AGN'' classification from \citet{2006MNRAS.372..961K} in the left panel of Fig.~\ref{fig:bpt}. The line ratios for \object{HE~2211-3903} lie on a narrow branch extending from \ion{H}{2} region-type to AGN-type line ratios. A significant number of pixels in \object{HE~2211-3903} belong to the composite classification. 
A strong contribution from LINER-type emission can be excluded based on the [\ion{S}{2}]/H$\alpha$ ratio. As shown in the right panel of Fig.~\ref{fig:bpt}, the few line ratios falling into the LINER region are still consistent with a non-LINER classification within their error bars.

The spatial location of the composite-type line ratios in \object{HE~2211-3903} reveals an anti-correlation with the \ion{H}{2} region complexes. The spatial assignment of line ratios classified as ``AGN'', ``\ion{H}{2}'' region, and ``composite'' in the left panel of Fig.~\ref{fig:bpt} is highlighted in Fig.~\ref{fig:compositeAGNHII}. Based on the [\ion{O}{3}]/H$\beta$ map, this plot shows pixels of ``AGN'', ``composite'', and ``\ion{H}{2}'' classification separately. \ion{H}{2}-type line ratios are associated with the \ion{H}{2} region complexes in the ring and spiral arms. Composite line ratios are found in the eastern half of the galaxy, predominantly tracing the low-density interstellar medium in the circum-nuclear region, in the bar-like structure, and between the \ion{H}{2} regions in the ring.

In the following sections, the three classes of emission-line ratios in \object{HE~2211-3903} are compared to existing photoionization models. The aim of this comparison is to use available photoionization models as a guidance for the line ratio interpretation. Given that the available models will not sample the full parameter space of the gas conditions in \object{HE~2211-3903}, a perfect match between models and data is not intended.

\subsubsection{The role of shocks \label{sec:shocks}}

\begin{figure*}
\epsscale{1.0}
\plotone{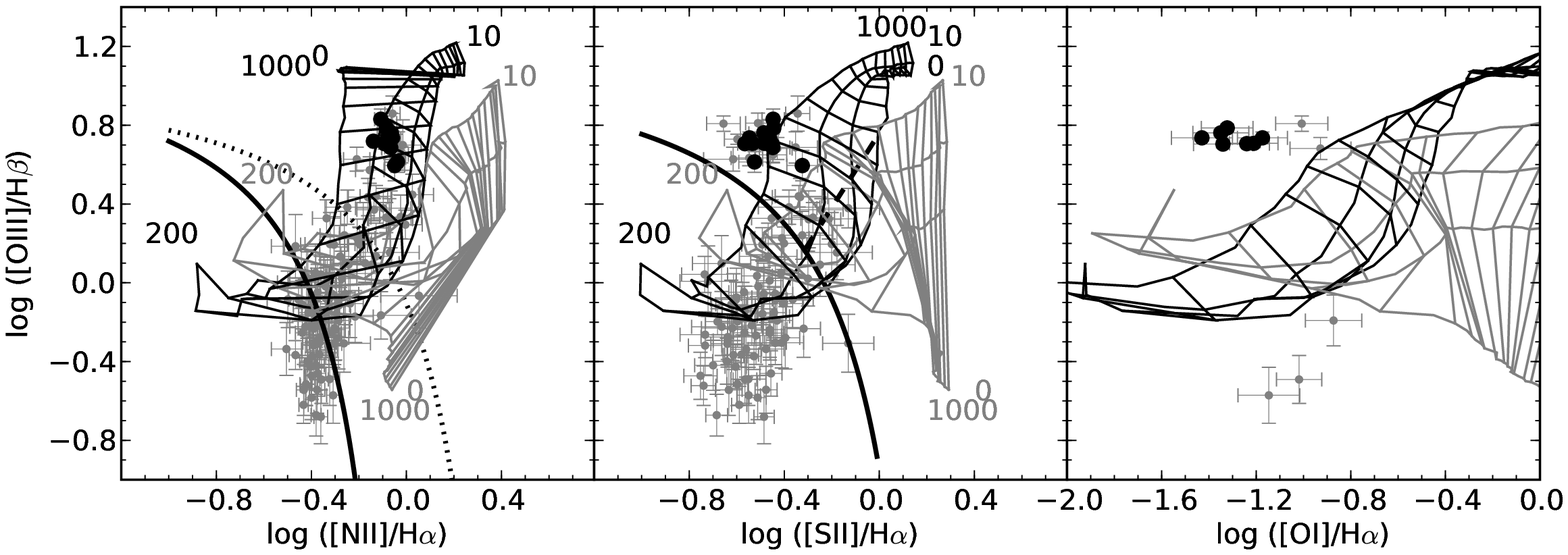}
\caption{[\ion{O}{3}]~$\lambda$5007/H$\beta$ versus [\ion{N}{2}]~$\lambda$6583/H$\alpha$, [\ion{S}{2}]~$\lambda$6717,6731/H$\alpha$, and [\ion{O}{1}]~$\lambda$6300/H$\alpha$ diagrams for HE~2211-3903 compared to a grid of radiative shock models taken from \citet{2008ApJS..178...20A}. Line ratios from within a radius of 2\arcsec\ around the AGN are marked in bold. The models shown here are the ``R'' models from \citet{2008ApJS..178...20A} for twice solar abundance and a
pre-shock density of $1\ \mathrm{cm^{-3}}$. The pure shock models and shock+precursor models are shown in gray and black, respectively. The shock velocities range from $200\ \mathrm{km\ s^{-1}}$ to $1000\ \mathrm{km\ s^{-1}}$ and are marked in steps of $50\ \mathrm{km\ s^{-1}}$. The lines of constant magnetic field are shown for $10^{-4}\ \mathrm{\mu G}$, $2\ \mathrm{\mu G}$, $4\ \mathrm{\mu G}$, and $10\ \mathrm{\mu G}$. Like in Fig.~\ref{fig:bpt}, the solid, dotted, and dashed lines outline the borders between the loci occupied by \ion{H}{2}, composite, Seyfert, and LINER line ratios. \label{fig:bptshocks}}
\end{figure*}
\begin{figure}
\epsscale{1.0}
\plotone{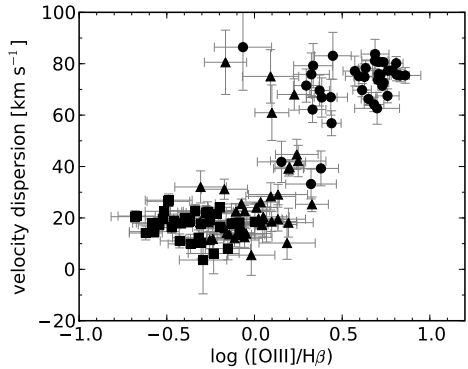}
\caption{Relation between the velocity dispersion and the [\ion{O}{3}]~$\lambda$5007/H$\beta$ ratio. AGN-type, composite, and \ion{H}{2} region-type line ratios are marked with circles, triangles, and squares, respectively.\label{fig:OIIIHbvdisp}}
\end{figure}
The WiFeS data do not show any conclusive evidence for ionization by fast radiative shocks in \object{HE~2211-3903}. In order to assess the role of fast shocks, the diagnostic diagrams for [\ion{N}{2}]/H$\alpha$ , [\ion{S}{2}]/H$\alpha$, and [\ion{O}{1}]/H$\alpha$ versus [\ion{O}{3}]/H$\beta$ are compared to the fast radiative shock models from \citet{2008ApJS..178...20A} in Fig.~\ref{fig:bptshocks}. It is obvious that a fraction of the composite and AGN-type ratios overlap with most of the shock+precursor models. However, two lines of argument suggest that this overlap is rather coincidental and not indicative of contribution from fast shocks.
The first argument is based on small velocity dispersions and the lack of a continuous transition from the higher velocity dispersions in the nuclear region to lower velocity dispersions in the disk. Fig.~\ref{fig:OIIIHbvdisp} or the right panel in Fig.~\ref{fig:kinematics} show that the maximum velocity dispersion in \object{HE~2211-3903} is about $80\ \mathrm{km\ s^{-1}}$. The regime of the fast shocks, however, corresponds to expected gas velocity dispersions of $\gtrsim 115\ \mathrm{km\ s^{-1}}$. This number is derived by dividing the fast shock model velocities of $\gtrsim 200\ \mathrm{km\ s^{-1}}$ by $\sqrt{3}$ in order to project the three-dimensional velocities onto the observed line-of-sight. In addition to the slow gas velocities in \object{HE~2211-3903}, Fig.~\ref{fig:OIIIHbvdisp} gives evidence of a discontinuity between the high velocity dispersions (around $80\ \mathrm{km\ s^{-1}}$) for predominantly AGN-type line ratios in the nucleus and the low velocity dispersions (around $20\ \mathrm{km\ s^{-1}}$) for mostly composite and \ion{H}{2} region-type ratios in the ring and spiral arms. This suggests that there is no large-scale evidence of AGN-driven shocks penetrating from the nuclear region into the composite and \ion{H}{2} regions. 
The second argument against shock-ionization in \object{HE~2211-3903} is based on the disagreement of the nuclear line ratios (data points marked in bold in Fig.~\ref{fig:bptshocks}) with the fast shock models.
Fig.~\ref{fig:bptshocks} shows that these nuclear line ratios are not represented well by the models in [\ion{S}{2}]/H$\alpha$ and [\ion{O}{1}]/H$\alpha$ versus [\ion{O}{3}]/H$\beta$. (The [\ion{N}{2}]/H$\alpha$ ratios are disregarded here because of the effects of a nuclear N-overabundance, as discussed in Section~\ref{sec:AGN}.)  
This suggests that large-scale AGN-driven shocks are unlikely to be significant in the nuclear region and, consequently, in any off-nuclear region. Shocks on scales smaller than the WiFeS resolution could remain undetected.

\subsubsection{\ion{H}{2} regions \label{sec:HII}}

\begin{figure*}
\epsscale{0.7}
\plotone{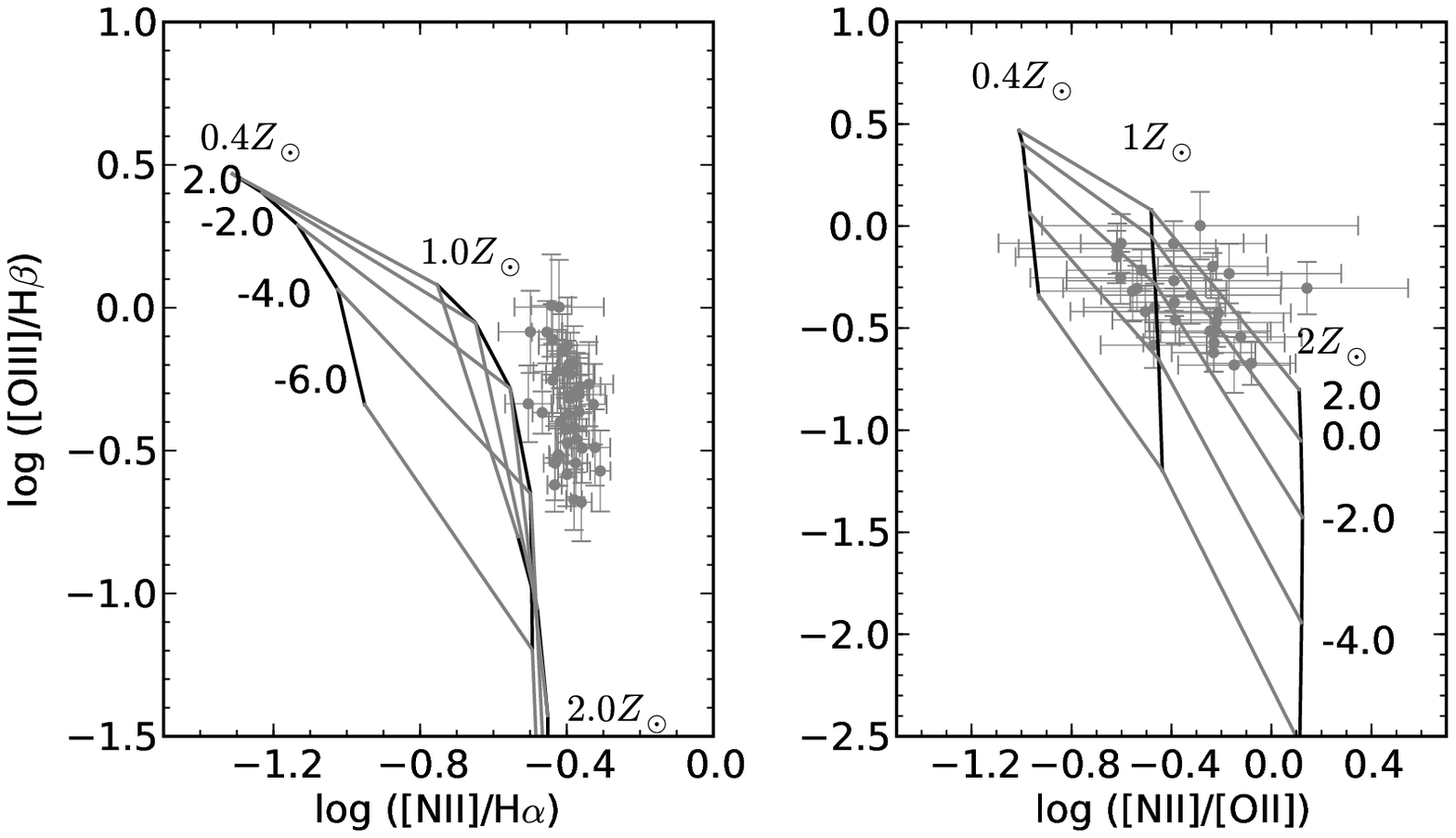}
\caption{Comparison of the diagnostic diagrams for HE~2211-3903 with photoionization models for ensembles of evolving \ion{H}{2} regions from \citet{2006ApJS..167..177D}. The model grids show varying metallicity of $Z=0.4\ Z_\odot$, $1.0\ Z_\odot$, $2.0\ Z_\odot$ and $\log R= 2$, 0, -2, -4, -6 \citep[for details see][]{2006ApJS..167..177D}. The observational data only include pixels with line ratios indicative of \ion{H}{2} regions based on their [\ion{N}{2}]/H$\alpha$ versus [\ion{O}{3}]/H$\beta$ diagnostics in Fig.~\ref{fig:bpt}. The [\ion{N}{2}]/[\ion{O}{2}] ratios are based on extinction-corrected line fluxes.\label{fig:bptHIImodels}}
\end{figure*}
The \ion{H}{2} region-type emission-line regions in \object{HE~2211-3903} indicate gas-phase metallicities of $1\ Z_\odot$ to $2\ Z_\odot$.
This is derived from comparing \ion{H}{2}-type line ratios with the photoionization models for ensembles of evolving \ion{H}{2} regions from \citet{2006ApJS..167..177D} in Fig.~\ref{fig:bptHIImodels}.
The left-hand diagram shows the classical \citet{1987ApJS...63..295V} diagnostic of [\ion{N}{2}]/H$\alpha$ versus [\ion{O}{3}]/H$\beta$. The right-hand panel shows the strongly abundance-sensitive diagnostic of [\ion{N}{2}]/[\ion{O}{2}] versus [\ion{O}{3}]/H$\beta$ from \citet{2000ApJ...542..224D}. The [\ion{N}{2}]/[\ion{O}{2}] ratios are extinction-corrected. This and the fact that [\ion{O}{2}] is located in a lower S/N part of the spectrum contribute to the larger error bars in this ratio. 
As described in \citet{2006ApJS..167..177D}, the models are parametrized by metallicity and $\log R$, where $R$ is defined as $(M_{cl}/M_\odot)/(P_0/k)$, i.e. the mass of the central cluster $M_{cl}$, measured in solar masses, divided by the ratio of the pressure of the interstellar medium $P_0$ and the Boltzmann constant $k$, measured in units of $\mathrm{cm^{-3}\ K}$.
Within the error bars, most data points of the abundance-sensitive ratio [\ion{N}{2}]/[\ion{O}{2}] agree with the models for metallicities of about $1\ Z_\odot$ to $1.5\ Z_\odot$ for $\log R >-4.0$. Based on typical cluster masses and pressures, \citet{2006ApJS..167..177D} expect such values of $\log R$, in particular $\log R =-2.0$ for disk galaxies. 
[\ion{N}{2}]/H$\alpha$ versus [\ion{O}{3}]/H$\beta$ becomes less abundance-sensitive at the high metallicity end, where metallicities of $1\ Z_\odot$ and $2\ Z_\odot$ produce a rather sharp cut-off at 
$\log$([\ion{N}{2}]/H$\alpha$)$\approx -0.5$. The line ratios measured for the \ion{H}{2} regions in \object{HE~2211-3903} lie close to the models for metallicities of $1\ Z_\odot$ and $2\ Z_\odot$ and 
$\log R > -4.0$. But all data points are offset by about 0.1~dex to 0.2~dex towards higher [\ion{N}{2}]/H$\alpha$ ratios. This indicates a systematic offset in the models or the measurements. 
The models can be affected by uncertainties in the N abundance which is strongly dependent on metallicity in metal-rich environments where N becomes a secondary element. The treatment of the N-abundance in the models is explained in detail in \citet{2006ApJS..167..177D} and \citet{2004ApJS..153....9G}. While the model N-abundance is a possible bias, it is unlikely to be the main explanation for the offset found in Fig.~\ref{fig:bptHIImodels}. An underestimation of the N-abundance would affect [\ion{N}{2}]/H$\alpha$ and [\ion{N}{2}]/[\ion{O}{2}] in a similar manner, which is not reflected in the data. One possibility is that the large [\ion{N}{2}]/H$\alpha$ measurements are caused by an underestimate of the contribution from H$\alpha$ absorption during the subtraction of the stellar continuum. However, 0.1~dex to 0.2~dex corresponds to flux factors of 1.3 to 1.5, which seems a large correction when attributing it to H$\alpha$ absorption only. Another possibility is that the \ion{H}{2} region line ratios are still affected by a residual contribution from AGN photoionization. A pre-absorbed AGN continuum can preferentially increase the contribution from low-ionization lines. If [\ion{O}{2}] and [\ion{N}{2}] are similarly affected, then the effect cancels out in the line ratio [\ion{N}{2}]/[\ion{O}{2}] but remains visible in [\ion{N}{2}]/H$\alpha$. As discussed in Section~\ref{sec:composite}, composite line ratios are found in between the \ion{H}{2} regions in the ring, which suggests that also the \ion{H}{2} regions can be affected by a contribution from residual AGN photoionization.

\subsubsection{AGN photoionization regions \label{sec:AGN}}

\begin{figure*}
\epsscale{1.0}
\plotone{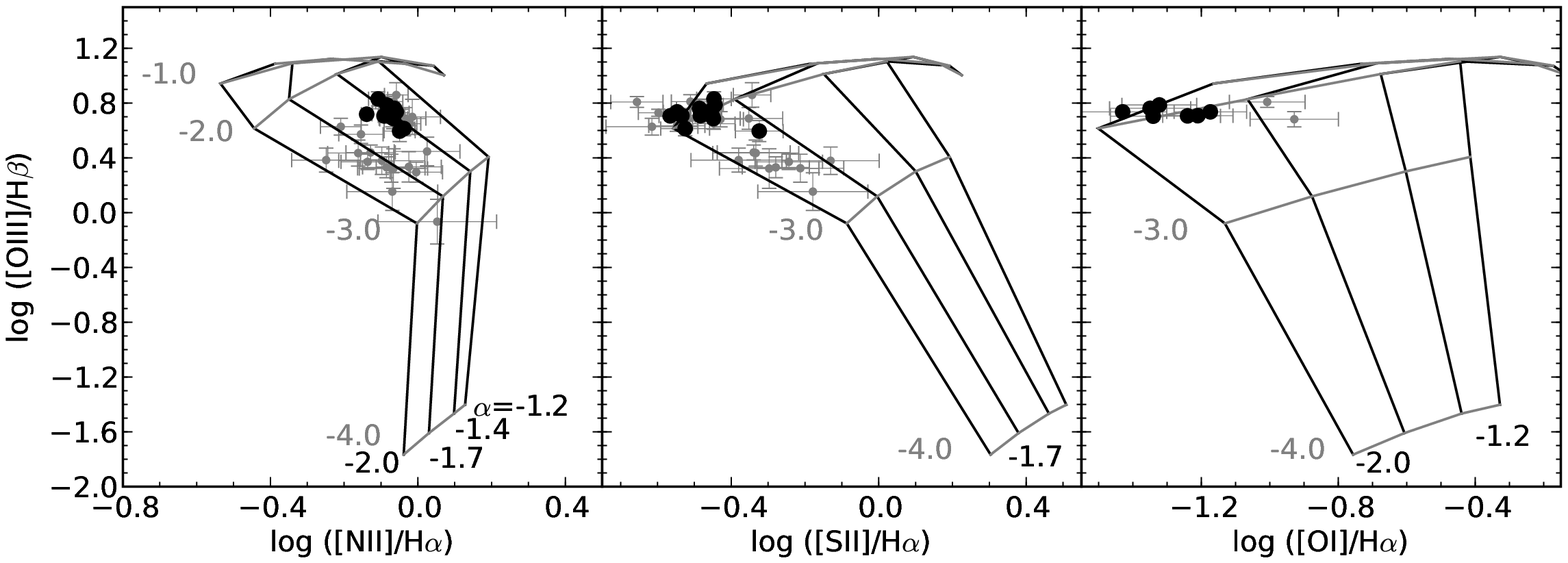}
\caption{Comparison of the diagnostic diagrams for HE~2211-3903 with dusty radiation-pressure-dominated photoionization models for the NLR from \citet{2004ApJS..153....9G}. Only the pixels with AGN-type line ratios corresponding to the left panel in Fig.~\ref{fig:bpt} are plotted. All data points are from within a radius of about 4\arcsec\ from the nucleus. Data points located within a 2\arcsec\ radius around the AGN are marked in bold. The model grid from \citet{2004ApJS..153....9G} shows the variation of the emission-line ratios with varying external ionization parameter and power-law index of the continuum source for an assumed metallicity of $Z = 2.0\ Z_\odot$ and a hydrogen density of $n_H=1000\ \mathrm{cm^{-3}}$. Ionization parameters and power-law indices are labeled.\label{fig:bptAGNmodels1}}
\end{figure*}
\begin{figure*}
\epsscale{0.7}
\plotone{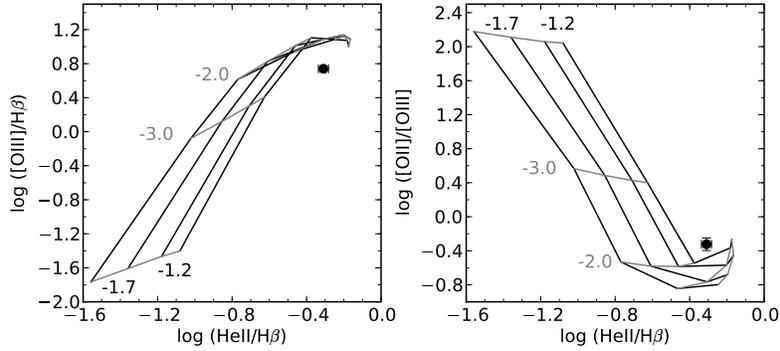}
\caption{Comparison of the nuclear line ratios of \ion{He}{2}~$\lambda$4686/H$\beta$ versus [\ion{O}{3}]~$\lambda 5007$/H$\beta$ and [\ion{O}{2}]~$\lambda\lambda 3727,29$/[\ion{O}{3}] with the dusty radiation-pressure-dominated photoionization models for the NLR from \citet{2004ApJS..153....9G}. The line ratios are measured in the nuclear spectrum integrated over a 2\arcsec\ aperture. The line fluxes are based on the fit of Gaussian profiles to the lines, as described in Section~\ref{sec:linefitting}, with the exception that, unlike the other lines, the width of the single Gaussian component for \ion{He}{2} is not tied to the width of the narrow H$\beta$ component. The model grid from \citet{2004ApJS..153....9G} shows the variation of the emission-line ratios with varying external ionization parameter and power-law index of the continuum source for an assumed metallicity of $Z = 2.0\ Z_\odot$ and a hydrogen density of $n_H=1000\ \mathrm{cm^{-3}}$. 
The ionization parameter is plotted for $U = 0.0$, -1.0, -2.0, -3.0, and -4.0. The power-law index is plotted for $\alpha = -1.2$, -1.4, -1.7, and -2.0.\label{fig:bptAGNmodels2}}
\end{figure*}
\begin{figure}
\epsscale{1.0}
\plotone{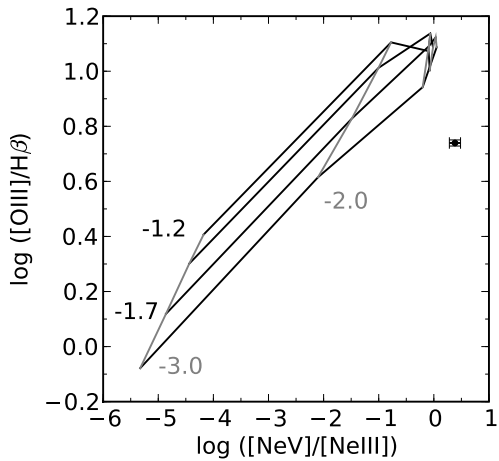}
\caption{Comparison of [\ion{Ne}{5}]~$\lambda$3426/[\ion{Ne}{3}]~$\lambda$3869 versus [\ion{O}{3}]~$\lambda 5007$/H$\beta$ with the dusty radiation-pressure-dominated photoionization models for the NLR from \citet{2004ApJS..153....9G}. The line ratios are measured in the nuclear spectrum integrated over a 2\arcsec\ aperture. Unlike other errors in this paper, the error bar for [\ion{Ne}{5}]/[\ion{Ne}{3}] is based on the range of line fluxes obtained by different methods of line integrations (i.e. Gaussian profile fits and direct line integrations). The model grid from \citet{2004ApJS..153....9G} shows the variation of the emission-line ratios with varying external ionization parameter and power-law index of the continuum source for an assumed metallicity of $Z = 2.0\ Z_\odot$ and a hydrogen density of $n_H=1000\ \mathrm{cm^{-3}}$. 
The ionization parameter is plotted for $U = 0.0$, -1.0, -2.0, and -3.0. The power-law index is plotted for $\alpha = -1.2$, -1.4, -1.7, and -2.0. \label{fig:bptAGNmodels3}}
\end{figure}

The comparison of the AGN-type emission-line gas in \object{HE~2211-3903} with available AGN photoionization models from \citet{2004ApJS..153....9G} suggests a N-overabundance in the nuclear region. In addition, the comparison provides an estimate of the power-law index of the ionizing AGN continuum as well as evidence of decreasing ionization parameter with increasing radial distance from the AGN. 
Diagnostic diagrams for all AGN-classified line ratios from the nuclear and circum-nuclear region or for single line ratios from an integrated nuclear spectrum are plotted in Figs.~\ref{fig:bptAGNmodels1}, \ref{fig:bptAGNmodels2}, and \ref{fig:bptAGNmodels3}. The model grids show solutions from the dusty, radiation-pressure-dominated photoionization models of \citet{2004ApJS..153....9G} for a variation of AGN power-law indices and external ionization parameters at a fixed metallicity of $2\ Z_\odot$. The models are available for a hydrogen density of $n_H= 1000\ \mathrm{cm^{-3}}$. The model solar abundance set and the scaling applied is described in detail in \citet{2004ApJS..153....9G}. \citet{2004ApJS..153....9G} define the ionization parameter
 as the ratio between the number of ionizing photons per unit area per time $S_\star$ and the hydrogen density $n_H$ times the speed of light $c$: $U = S_\star / n_Hc$. The AGN power-law continuum is described between a minimum and maximum frequency of 5~eV and 1000~eV, respectively, as $F_\nu \propto \nu^\alpha$.
The assumption of a metallicity of $2\ Z_\odot$ is based on an extrapolation from the gas-phase metallicity in the 
\ion{H}{2} regions of $1\ Z_\odot$ to $2\ Z_\odot$, as discussed in Section~\ref{sec:HII}. The assumption of twice solar metallicity models for the nuclear region of \object{HE~2211-3903} is furthermore supported by the fact that the models from \citet{2004ApJS..153....9G} for $1\ Z_\odot$ do not reproduce well the majority of the measured line ratios.

First, the results from the three diagnostic diagrams in Fig.~\ref{fig:bptAGNmodels1} are discussed.
The highest [\ion{O}{3}]/H$\beta$ ratios are found in the nucleus of \object{HE~2211-3903} (points marked in bold), which cluster in a narrow region in the diagnostic diagrams. The comparison of all three diagnostic diagrams with the model grids shows that the nuclear [\ion{N}{2}]/H$\alpha$ ratios deviate from the [\ion{S}{2}]/H$\alpha$ and [\ion{O}{1}]/H$\alpha$ ratios in terms of their model interpretation: Both, [\ion{S}{2}]/H$\alpha$ and [\ion{O}{1}]/H$\alpha$ versus [\ion{O}{3}]/H$\beta$, agree with the dusty radiation-pressure-dominated models for a power-law index close to -2.0 and an ionization parameter of $\log U \approx -1.9$.  
With respect to this, [\ion{N}{2}]/H$\alpha$ is offset horizontally by +0.4~dex to a power-law index of -1.4 and an ionization parameter between $\log U \approx -2.0$ and $\log U \approx -3.0$. This 
deviation of the nuclear  [\ion{N}{2}]/H$\alpha$ ratios can be explained via a nuclear overabundance of N in \object{HE~2211-3903}. An indication of N-overabundance also remains when taking into account a possible intrinsic offset between the [\ion{N}{2}]/H$\alpha$ measurements and the model values of the order of 0.1~dex to 0.2~dex, as found for the \ion{H}{2} regions in Section~\ref{sec:HII}. The scenario of N-overabundance is supported by the fact that the offset is predominantly confined to the nucleus itself. At a distance of 2\arcsec\ from the nucleus, where the [\ion{O}{3}]/H$\beta$ ratios start to decrease, the line ratios occupy a similar model parameter space in [\ion{N}{2}]/H$\alpha$ and [\ion{S}{2}]/H$\alpha$. These circum-nuclear line ratios are, therefore, in agreement with a normal N-abundance. For the circum-nuclear region,  the [\ion{S}{2}]/H$\alpha$ versus [\ion{O}{3}]/H$\beta$ diagram suggests a trend towards lower ionization parameters at constant power-law index with increasing distance from the AGN. This is in agreement with a scenario in which the gas in the circum-nuclear region is affected by a lower ionizing flux from the same AGN continuum source. Outside the inner 2\arcsec\ around the AGN, the ionization parameter is $\log U < -2.0$ and it decreases to $\log U \approx -2.8$ at 3\arcsec\ from the nucleus. 
An alternative estimate of the latter ionization parameter, mostly independent of the shape of the ionizing continuum, can be directly obtained from
[\ion{O}{2}]~$\lambda\lambda 3727,29$/[\ion{O}{3}]~$\lambda 5007$ \citep[see][]{1997A&A...323...31K}. In the spectrum integrated over a 2\arcsec\ nuclear aperture on \object{HE~2211-3903}, this ratio is about -0.3 (right panel in Fig.~\ref{fig:bptAGNmodels2}). Compared to Fig.~7 in \citet{1997A&A...323...31K}, this results again in an ionization parameter of about $\log U = -2.7$ when assuming a low density of 
$n_H \lesssim 10^{3}\ \mathrm{cm^{-3}}$.
The decrease in ionization parameter from the nucleus to a distance of about 3\arcsec\ is in agreement with a decrease proportional to the inverse square of the distance from the ionizing source, as expected in a constant-density environment.

The following calculation shows that the flux of ionizing photons from the AGN required for the derived ionization parameter at about 3\arcsec\ from the AGN is consistent with the intrinsic luminosity of the AGN from Section~\ref{sec:AGNpower}. The photon flux is $S_\star = U \times  n_H c$. Assuming a density of $n_H = 100\ \mathrm{cm^{-3}}$ (Section~\ref{sec:electrondensity}) and an ionization parameter of $\log U \approx -2.8$, the photon flux from the AGN at 3\arcsec, or 2.3~kpc, distance is $S_\star \approx 5\times 10^{9}\ \mathrm{cm^{-2}\ s^{-1}}$. For a photon energy of $E_\star = 13.6\ \mathrm{eV}$, the corresponding energy flux is $F_\star = S_\star \times E_\star = 0.1\ \mathrm{erg\ cm^{-2}\ s^{-1}}$. If this energy flux is radiated isotropically by the AGN into distances of $R=2.3$~kpc, the AGN luminosity is $L_\star = 4\pi R^2 \times F_\star = 7\times 10^{43}\ \mathrm{erg\ s^{-1}}$. This value is well within the limit given by the bolometric luminosity of the AGN derived in Section~\ref{sec:AGNpower}.

In view of the hypothesis of nuclear N-enrichment, the three additional line ratios involving \ion{He}{2}~$\lambda$4686, [\ion{Ne}{5}]~$\lambda$3426, and [\ion{Ne}{3}]~$\lambda$3869 are compared to the dusty radiation-pressure-dominated models in Figs~\ref{fig:bptAGNmodels2} and \ref{fig:bptAGNmodels3}. These lines can only be measured at sufficient S/N in the nuclear spectrum. 
The width of the \ion{He}{2}~$\lambda$4686 line in the nuclear spectrum is broader by a factor of about 1.5 than the width of the other narrow lines. This is taken into account during the fit to the nuclear spectrum, by decoupling the width of \ion{He}{2} from the H$\beta$ line. Like [\ion{N}{2}]/H$\alpha$, \ion{He}{2}/H$\beta$ shows an offset of about +0.4~dex compared to the line ratios predicted by the models for an AGN power-law index of -2.0. This suggests that the N-enrichment is accompanied by He-enrichment. 
[\ion{Ne}{5}]/[\ion{Ne}{3}] (Fig.~\ref{fig:bptAGNmodels3}) is the only abundance-independent line ratio. Unlike the other line ratios, [\ion{Ne}{5}]/[\ion{Ne}{3}] deviates significantly from the model parameter space. [\ion{Ne}{5}] is located in a low S/N region of the blue spectrum close to the band limit. In this region, flat fielding becomes unreliable and systematic errors might be introduced in addition to the statistical errors shown in Fig.~\ref{fig:bptAGNmodels3}. Furthermore, the deviation between the data and the model could reflect the known tendency of photoionization models to underpredict the strongest observed [\ion{Ne}{5}]/[\ion{Ne}{3}] ratios \citep[e.g.][]{1996A&A...312..365B, 1997A&A...323...31K, 1997ApJ...487..122F}. It is possible that the bulk of the [\ion{Ne}{5}] emission of \object{HE~2211-3903} arises at very small nuclear distances at high density and in perhaps dust-free environments, where the models of \citet{2004ApJS..153....9G} are no longer fully applicable.

Dilution with nuclear star formation can produce AGN-type line ratios which suggest an artificially softer AGN power-law. In particular, the single Gaussian line profile for the narrow component does not fully account for the complexity of the narrow lines in the nuclear region. This is a possible uncertainty in the model interpretation of the measured AGN-type line ratios presented here.

\subsubsection{Regions with composite line ratios\label{sec:composite}}

\begin{figure}
\epsscale{0.9}
\plotone{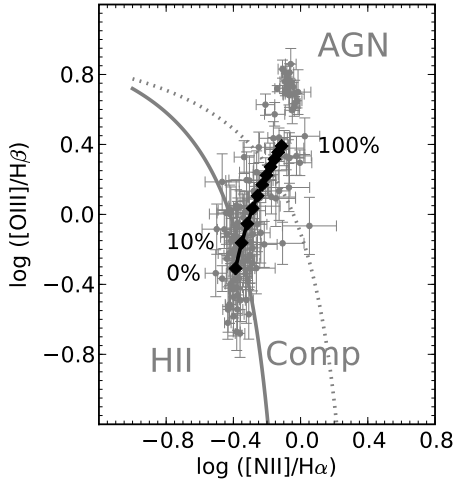}
\caption{[\ion{O}{3}]~$\lambda$5007/H$\beta$ versus [\ion{N}{2}]~$\lambda$6583/H$\alpha$ diagram for HE~2211-3903 (grey error bars) compared to a mixing curve of photoionization from the AGN and from star formation (black curve). The percentage of AGN contribution to the \ion{H}{2} region values is shown in steps of 10\% along the mixing curve (black diamonds). As in Fig.~\ref{fig:bpt}, the gray solid and dotted curves outline the borders between the loci occupied by AGN-type, composite, and \ion{H}{2} region-type line ratios, based on \citet{2006MNRAS.372..961K}. The anchor points of the mixing curve are explained in the text.\label{fig:bptmixing}}
\end{figure}

The composite line ratios in the circum-nuclear region and the low-density interstellar medium between the \ion{H}{2} regions in the eastern part of the bar-like structure and the ring of \object{HE~2211-3903} (Fig.~\ref{fig:compositeAGNHII}) are consistent with a varying degree of mixing between photoionization purely by the AGN and purely by star formation along a theoretical mixing curve. 
Such a mixing line has been discussed in the context of diagnostic diagrams for a sample of integrated galaxy emission-line ratios \citep[][]{2001ApJS..132...37K}.
Fig.~\ref{fig:bptmixing} shows that such a mixing curve likewise reproduces the different types of spatially resolved composite line ratios in \object{HE~2211-3903}. 
The anchor points of the mixing curve are estimated according to the following criteria:
The anchor point for pure AGN photoionization is chosen as the average of all off-nuclear AGN-type line ratios at $\log U \approx -2.8$ (see Fig.~\ref{fig:bptAGNmodels1}).
This ionization parameter is likely to be most representative of the conditions of AGN photoionization further out in the galaxy. The anchor point for pure \ion{H}{2} region photoionization is selected to be the average value of all line ratios classified as \ion{H}{2}-region-like according to \citet{2006MNRAS.372..961K}.
In terms of this hypothetical mixing line, the composite line ratios in \object{HE~2211-3903} agree with AGN contributions to the photoionization between about 10\% to 70\%. On average [\ion{O}{3}]~$\lambda$5007/H$\beta$ for the composite ratios is large close to the AGN and decreases radially into the outer parts of the ring (Fig.~\ref{fig:allratios} or the color version of Fig.~\ref{fig:compositeAGNHII}). This agrees with a scenario of decreasing AGN contribution to the photoionization with increasing distance from the AGN.

\section{Discussion}

\object{HE~2211-3903} was chosen from a sample of Seyfert/QSO-borderline type-1 AGN in order to probe the properties of a bright AGN and its host galaxy at low redshift via spatially resolved emission-line diagnostics.

A basic morphological assessment of the emission-line data presented in this paper suggests that the AGN of \object{HE~2211-3903} resides in a barred ringed galaxy. Regarding the overall regular velocity field, it is likely that \object{HE~2211-3903} is currently dominated by secular rather than merger-driven evolution. Secular evolution is considered to play an important role in fueling AGN at low redshift, while powerful high-redshift QSOs are more likely to be driven by major galaxy mergers \citep[e.g.][and references therein]{2011ApJ...726...57C}. In this scenario \object{HE~2211-3903} ranks among typical low-redshift AGN.

The emission-line data for \object{HE~2211-3903} show signs of N-overabundance which is confined to the nucleus. AGN with N-overabundance have been found from low to high redshift. E.g. \citet{1999ARA&A..37..487H} review the case of high relative N-abundance for high-redshift QSOs. In the case of low-redshift AGN,                                                                                                                                                                                                                            \citet{1991MNRAS.249..404S} concludes that the majority of objects from a sample of Seyfert~2 and LINER galaxies show evidence of N-overabundance in the nucleus. Similarly, by comparison with fast shock models, \citet{1995ApJ...455..468D} find N-overabundance for Seyfert 1.5, Seyfert 2, and LINER galaxies. They also demonstrate that increasing the input N-abundance in the models directly translates into an increase in [\ion{N}{2}]/H$\alpha$. The single integrated spectrophotometric data in their study show N-overabundance of up to 0.4~dex. This range is similar to the value derived for \object{HE~2211-3903}. The spatially resolved information for \object{HE~2211-3903} reveals that the N-overabundance is associated with the nucleus, where the interstellar medium must have been chemically enriched during past and/or ongoing nuclear starburst activity. The finding that He appears to be enriched together with N is not unexpected. The stars producing N enrichment at the same time eject He which is added to the existing primordial He abundance.
The main source for chemical enrichment of the interstellar medium is mass loss from Asymptotic Giant Branch (AGB) stars, Wolf-Rayet stars, and supernovae \citep[e.g.][]{2011arXiv1102.5312K}.
Mass loss from AGB stars mainly contributes to the enrichment of He, C, and N. Mass loss from Wolf-Rayet stars enriches the interstellar medium with He and N as well as, in the later stages of evolution, C and O. Assuming initial mass functions for the stellar populations, \citet{2003MNRAS.338..973D} show that, at solar metallicity, AGB stars of $\sim 4-8\ M_\odot$ play a dominant role in N-enrichment, while the C- and O-enrichment is dominated by Wolf-Rayet stars and supernovae, respectively. The detailed share of AGB and Wolf-Rayet stars in their N-enrichment scenarios depends on the initial mass function.
The typical life time of AGB stars is $\sim 10^7$~yrs, while that of Wolf-Rayet stars is a few $10^6$~yrs. When attributing the N-enrichment to AGB stars in a single starburst only, this means that the starburst must be at least $\sim 10^7$~yrs old. Similarly, in the case of N-enrichment by Wolf Rayet stars, the starburst must be at least a few $10^6$~yrs old.
Based on the analysis presented here, it is not possible to rule out N-enrichment from starbursts further in the past, nor to exclude the scenario of multiple successive starburst episodes.

The infrared brightness of \object{HE~2211-3903} supports a scenario of recent and ongoing starburst activity. In particular, the IRAS colors of $\alpha(100,60)=-0.58$ and $\alpha(60,25) = -0.61$ fall into a region of the IRAS two-color diagram which is between that of black-body-dominated luminous infrared galaxies and power-law-dominated QSOs and Seyfert galaxies \citep{1994ApJ...436..102L, 2001ApJ...555..719C}. Objects in this region of the two-color diagram have been interpreted as ``transition objects'' in a possible evolutionary sequence from an ultra-luminous infrared to an AGN phase \citep{2001ApJ...555..719C}. This evolutionary sequence is a scenario for the merger-driven evolution of sources with $L_{IR} > 10^{12}\ L_\odot$ evolving into luminous optical QSO nuclei. Whether such an interpretation as a transition object can be directly transferred to low-redshift AGN like \object{HE~2211-3903} with scaled-down luminosities and dominated by secular evolution remains to be investigated.

For the classical transition objects \citet{2001ApJ...555..719C} find a time lag of about $50\times 10^6$~yrs between the peak of the starburst activity and the radio-quiet AGN becoming visible. This is a similar delay time as the $50\times 10^6$~yrs to $100 \times 10^6$~yrs found by \citet{2007ApJ...671.1388D} for the minimum age of the star formation in efficiently fueled AGN in nearby Seyfert galaxies. \citet{2007ApJ...671.1388D} discuss the role of winds and outflows from different types of stars in hindering or supporting AGN accretion. They argue that the slow winds from AGB stars play an important role for efficient accretion of the ejecta onto the AGN in contrast to the fast winds from Wolf-Rayet stars and supernovae which rather prevent efficient accretion.
If this scenario is applied to \object{HE~2211-3903}, the slow winds from AGB stars formed in a starburst about $\sim 10^7$~yrs ago can be the origin of both, the observed N-enrichment and the fueling of the currently optically bright AGN.

The AGN in \object{HE~2211-3903} in turn has an observable effect on the host galaxy. The composite line ratios are likely to trace a contribution of AGN photoionization of the low-density interstellar medium out to distances of about 8~kpc from the AGN. 
With the wide azimuthal coverage and without signs of shock-excitation, the properties of this ENLR in \object{HE~2211-3903} are different from those of many Seyfert 1.5 and 2 galaxies in the literature.
The emission-line region in \object{HE~2211-3903} covers an angle of almost 180~deg in the eastern half of the ring. This wide opening angle is distinct from the elongated extended narrow-line emission typical of Seyfert 1.5 and 2 galaxies \citep[e.g.][]{1987MNRAS.228..671U}. Various studies show that the elongations of the latter are usually correlated with the radio morphology \citep[e.g.][]{1987MNRAS.228..671U, 1994ApJ...421...87C, 1997ApJ...487L.121W}. \citet{1995ApJ...455..468D} and \citet{1996ApJS..102..161D} suggested that the extended emission-line regions in Seyfert 1.5 and 2 galaxies are shock-excited via mechanical energy input from jets.
For \object{HE~2211-3903} the line ratios and velocity dispersions presented in Section~\ref{sec:shocks} show no evidence for shock-excitation of the gas in the ENLR. Furthermore, the wide opening angle of the ENLR over almost 180~deg in the ring supports a scenario without large-scale radio-jet at least in the plane of the ring. 
In available radio data from the NRAO VLA Sky Survey, the radio source of 3.8~mJy at 1.4~GHz is reported to be extended but not resolved into components \citep{2000yCat..21290547B}. The search for a possible radio jet could be the motivation for future radio observations of \object{HE~2211-3903}.

The analysis of the galaxy-scale EELR in \object{HE~2211-3903} suggests that, even without an obvious contribution from a radio jet, AGN photoionisation alone can have a subtle effect on the interstellar medium in host galaxies. In luminous quasars AGN photoionization has been found to affect the whole host galaxy, as discussed by \citet{2011ApJ...732....9G} based on long-slit spectroscopy of a sample of 15 obscured quasars at $0.1 < z < 0.5$. But detailed observations and analyses become increasingly difficult at higher redshift and AGN luminosity. Observations are affected by both, decreasing spatial resolution as well as contamination from the PSF of the bright nucleus. In a spatially unresolved spectrum integrated over the whole WiFeS field-of-view, \object{HE~2211-3903} would be classified as ``composite'' based on the narrow-line ratios. These ratios are $\log$~[\ion{N}{2}]/H$\alpha = -0.2$, $\log$~[\ion{O}{3}]/H$\beta =  0.2$, and $\log$~[\ion{S}{2}]/H$\alpha = -0.5$ when fitting the narrow lines with a single Gaussian profile. Such a spectrum only indicates a superposition of emission from the AGN NLR and from star formation throughout the host galaxy. But the spatial information is necessary in order to detect the off-nuclear composite emission as sign of an ENLR. By extending the sample of classical radio-loud and radio-quiet low-redshift QSOs with EELR detections \citep[e.g.][]{2008A&A...488..145H, 2009ApJ...690..953F} towards lower redshifts and AGN luminosities, objects like \object{HE~2211-3903} at the bright end of the $z<0.06$ Seyfert population are likely to provide further insights into the properties and physics of EELRs.

\section{Summary}

The spatially resolved emission-line gas properties of the bright Seyfert 1.5 galaxy \object{HE~2211-3903} were analyzed based on optical integral field spectroscopy using the Wide Field Spectrograph WiFeS.

A ring of ionized gas with a projected radius of about 6~kpc is revealed by the emission-line maps of \object{HE~2211-3903}. A linear structure of ionized gas connecting the ring with the nuclear region is likely to be associated with the stellar bar mentioned in previous literature. Based on the ring morphology as well as the global H$\alpha$ kinematics, a close to face-on view onto \object{HE~2211-3903} is suggested. The overall H$\alpha$ kinematics is indicative of disk rotation.
\object{HE~2211-3903} is tentatively discussed as an example of a barred ringed galaxy and as a candidate for AGN fueling by secular evolution, similar to many low-redshift AGN.

Emission-line diagnostics and a rather abrupt transition from narrow emission lines in the ring and spiral arms to broadened emission lines in the AGN NLR indicate that shock ionization is likely to be negligible at the spatial resolution of the observations. Instead, a scenario of AGN photoionization of the low-density interstellar medium on galaxy scales is discussed. This ENLR is traced via composite emission-line ratios on scales of up to 8~kpc. Emission from this ENLR is detected over an angle of about 180~deg in the eastern half of the galaxy, including the diffuse gas in the circum-nuclear region, in the bar-like structure, and between the \ion{H}{2} regions in the ring. The AGN contribution to the composite line ratios tracing the ENLR decreases with increasing radial distance from the AGN.

The nucleus of \object{HE~2211-3903} shows signs of N-overabundance, accompanied by He-overabundance. The N-overabundance is found to be confined to data points from the very nucleus, as the circum-nuclear region and the \ion{H}{2} regions in the disk agree with a normal N-abundance. The most likely source for this enrichment are winds from AGB and/or Wolf-Rayet stars during past and possibly ongoing nuclear starburst episodes. Ongoing starburst activity is supported by the infrared luminosity and colors.

\acknowledgments
The authors thank the referee for a thorough review of the paper and helpful comments and suggestions.
M.A. Dopita and J. Scharw\"achter acknowledge the continued support of the Australian Research Council (ARC) through Discovery projects DP0984657 and DP0664434. 
Fig.~\ref{fig:dss} is based on data from the Digitized Sky Survey.
The Digitized Sky Survey was produced at the Space Telescope Science Institute under U.S. Government grant NAG W-2166. The images of these surveys are based on photographic data obtained using the Oschin Schmidt Telescope on Palomar Mountain and the UK Schmidt Telescope. The plates were processed into the present compressed digital form with the permission of these institutions. 
The UK Schmidt Telescope was operated by the Royal Observatory Edinburgh, with funding from the UK Science and Engineering Research Council (later the UK Particle Physics and Astronomy Research Council), until 1988 June, and thereafter by the Anglo-Australian Observatory. The blue plates of the southern Sky Atlas and its Equatorial Extension (together known as the SERC-J), the near-IR plates (SERC-I), as well as the Equatorial Red (ER), and the Second Epoch [red] Survey (SES) were all taken with the UK Schmidt telescope at the AAO. 


\end{document}